# Strong-coupling multigap superconductivity in the Heusler compound $ScAu_2Al$

Jozef Kačmarčík[1], Zuzana Pribulová[1], Yevhen Petrenko[1,2], Gabriel Kuderowicz[3], Paweł Wójcik[4], Michał J. Winiarski[5], Szymon M. Królak[5], Filip Košuth[1], Pavol Szabó[1], Tomasz Klimczuk[5], Bartlomiej Wiendlocha[4], and Peter Samuely[1*]

*[1]Centre of Low Temperature Physics, Institute of Experimental Physics, Slovak Academy of Sciences, 04001 Košice, Slovakia*
*[2]Centre of Low Temperature Physics, Faculty of Science, P. J. Šafárik University, SK-04001 Košice, Slovakia*
*[3]Centre of Excellence ENSEMBLE3 Sp. z o. o., Wolczynska Str. 133, Warsaw 01-919, Poland*
*[4]AGH University of Krakow, Faculty of Physics and Applied Computer Science, Aleja Mickiewicza 30, 30-059 Krakow, Poland*
*[5]Faculty of Applied Physics and Mathematics and Advanced Material Center, Gdansk University of Technology, ul. Narutowicza 11/12, Gdańsk 80–233, Poland*

$ScAu_2Al$ is the Heusler superconductor with the highest reported transition temperature, $T_c$, close to 5 K. Here, we combine high-resolution ac calorimetry and electrical-transport measurements with first-principles-based Eliashberg calculations. The large specific-heat jump, $\Delta C/\gamma_n T_c = 2.2$, well above the weak-coupling BCS value of 1.43, together with an electron-phonon coupling constant $\lambda \approx 1.1$, establishes $ScAu_2Al$ as a strong-coupling superconductor. The electronic specific heat is described better by a two-gap α model, with $2\Delta_S/k_BT_c = 4.1$ and $2\Delta_L/k_BT_c = 4.8$, than by a single-gap model. The thermodynamic upper critical field follows a conventional WHH-like temperature dependence with $B_{c2}(0) \approx 0.18$ T, whereas the resistively determined critical field reaches values about three times larger and exhibits a pronounced positive curvature. We attribute the enhanced resistive field scale to superconductivity in disordered grain-boundary and interfacial regions. The experimental results are supported by calculations of the heat capacity and upper critical field within the Eliashberg formalism using Fermi-surface and electron-phonon parameters obtained from first-principles calculations.

## I. INTRODUCTION

Heusler compounds with the cubic $MnCu_2Al$-type structure ($Fm\bar{3}m$) form a broad family of intermetallic materials displaying a wide range of physical properties, including various magnetic orders, heavy-fermion behavior, and superconductivity [1]. More than 30 Heusler compounds are known to superconduct [2], with transition temperatures $T_c$ ranging approximately from 0.74 to 5 K. The Matthias rule, which associates enhanced $T_c$ with a valence-electron concentration close to seven electrons per atom [3], has been discussed in this context [1, 4]. However, non-superconducting Pauli paramagnets with the same valence-electron concentration are also known. In addition, $T_c$ can be sensitive to small changes in nominal composition and stoichiometry [5]. Correlations of $T_c$ with the lattice parameter, Debye temperature, density of states at the Fermi level, and electron-phonon coupling strength λ have been examined [1]; among these quantities, electron-phonon coupling appears to play the dominant role within a conventional phonon-mediated framework.[1]

$ScAu_2Al$ has the highest reported $T_c$ among Heusler superconductors, close to 5 K. Previous results concerning its superconducting properties are, however, not fully consistent. Bag et al. [6] reported a normalized specific-heat jump $\Delta C/\gamma_n T_c$ smaller than the weak-coupling BCS value of 1.43, while estimating a moderate electron-phonon coupling constant $\lambda \approx 0.8$. In contrast, first-principles calculations [7] predict phonon-mediated strong-coupling superconductivity with $\lambda = 1.25$ and a two-band superconducting state with two nodeless gaps on distinct Fermi-surface sheets, $2\Delta_S/k_BT_c = 4.1$ and $2\Delta_L/k_BT_c = 4.3$.

To clarify these discrepancies, we investigate superconductivity in $ScAu_2Al$ using highly sensitive ac calorimetry, electrical transport, and complementary STM tunneling measurements. We find a large specific-heat jump $\Delta C/\gamma_n T_c = 2.2$ and an electron-phonon coupling constant $\lambda =$

* samuely@saske.sk

1.1, demonstrating strong-coupling superconductivity. The electronic specific heat is described slightly better by a two-gap α-model, with $2\Delta_S/k_BT_c = 4.1$ and $2\Delta_L/k_BT_c = 4.8$, than by a single-gap strong-coupling model. Thermodynamic upper critical field follows a conventional WHH-like temperature dependence with $B_{c2}(0) \approx 0.18$ T, whereas the resistively determined critical field reaches 0.5–0.6 T and displays positive curvature, consistent with previous transport measurements [6]. We attribute this higher resistive field scale to superconductivity persisting in disordered grain-boundary and interfacial regions of the polycrystalline sample.

These findings are complemented by calculations of the superconducting thermodynamic properties, including the heat capacity and upper critical field, within Eliashberg strong-coupling theory using the Eliashberg spectral function and Fermi-surface properties obtained from first-principles calculations.

## II. EXPERIMENTAL DETAILS

Polycrystalline $ScAu_2Al$ was synthesized by arc melting high-purity elements: Sc (99.9%, Onyxmet, Poland), Au (99.9%, Mennica-Metale, Poland), and Al (99.999%, Onyxmet, Poland), under high-purity Zr-gettered argon. The button was flipped and remelted several times to improve homogeneity. After melting, the sample was wrapped in thin tantalum foil, sealed under vacuum in a fused-silica tube, annealed at 750 °C for 6 days, and slowly cooled to room temperature. We found that annealing followed by slow cooling is crucial for obtaining high-quality samples with a higher $T_c$ and sharp superconducting transitions in magnetization, heat capacity, and resistivity.

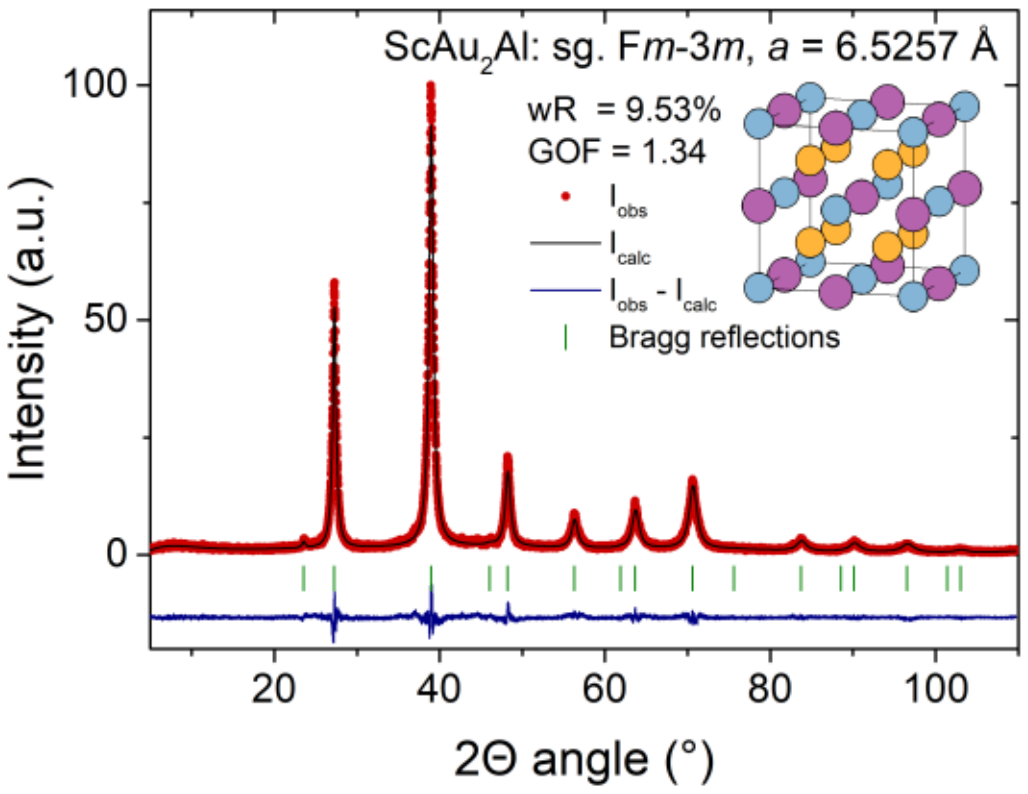

FIG. 1. PXRD pattern of annealed $ScAu_2Al$ (red points) with the Rietveld refinement (black line). Green ticks mark the Bragg-reflection positions of the full-Heusler structure; the unit cell is shown in the upper-right inset. The blue line is the difference between observed and calculated intensities.

The phase purity was assessed by powder x-ray diffraction (PXRD) using a Bruker D2 Phaser diffractometer with Cu Kα radiation and a LynxEye XE-T position-sensitive detector. All observed diffraction peaks were successfully indexed using the full-Heusler structure (space group $Fm\bar{3}m$, No. 225). Rietveld refinement performed with GSAS-II [8] yielded the lattice parameter $a$ = 6.5257(2) Å (Fig. 1).

For heat-capacity measurements, we first used the standard relaxation technique implemented in a Quantum Design Physical Property Measurement System (PPMS) down to 1.8 K to determine the absolute heat capacity. More detailed measurements were performed by ac calorimetry in a $^3$He refrigerator equipped with an 8 T superconducting magnet. ac calorimetry is particularly sensitive to small changes in the heat capacity of small samples and enables continuous measurements as a function of temperature and magnetic field [9, 10]. In one setup, the sample was heated at a modulation frequency of a few hertz using a light-emitting diode coupled to an optical fiber, and the resulting temperature oscillations were detected with a chromel-constantan thermocouple calibrated in magnetic field. We additionally used a newly developed resistive ac calorimeter based on a bare Cernox chip divided into separate heater and thermometer sections. This calorimeter reproduced the optical-heating data and provided an improved signal-to-noise ratio.

Electrical resistivity was measured using a standard four-probe technique in a $^3$He refrigerator equipped with an 8 T superconducting solenoid.

## III. EXPERIMENTAL RESULTS

Figure 2(a) shows the total specific heat of $ScAu_2Al$ in zero field (superconducting state) and at 1 T (normal state). A sharp superconducting transition begins slightly above 5 K, followed by a rapid decrease of the specific heat on cooling. The thermodynamic transition temperature $T_c$ = 4.95 K is determined by an equal-entropy construction. The superconducting specific heat divided by temperature $C_s(T)/T$ extrapolates to essentially zero as $T \rightarrow 0$, consistent with a fully superconducting sample and a negligible residual electronic contribution. The 1 T curve shows no superconducting anomaly down to the lowest measured temperature of 0.45 K and is therefore taken as the normal-state specific heat, $C_n(T)=\gamma_n T + C_{lattice}(T)$. Below approximately 0.9 K, $C_n/T$ becomes nearly temperature independent, allowing a reliable estimate of the Sommerfeld coefficient $\gamma_n \approx 10$ mJ mol$^{-1}$ K$^{-2}$.

The superconducting electronic contribution is obtained as $C_{es}/\gamma_n T = \Delta C/\gamma_n T + 1$ and is shown in Fig. 2(b). The superconducting anomaly is extremely sharp, indicating a high degree of superconducting homogeneity. Its normalized height, $\Delta C/\gamma_n T_c = 2.2$, is substantially larger than the weak-coupling BCS value of 1.43.

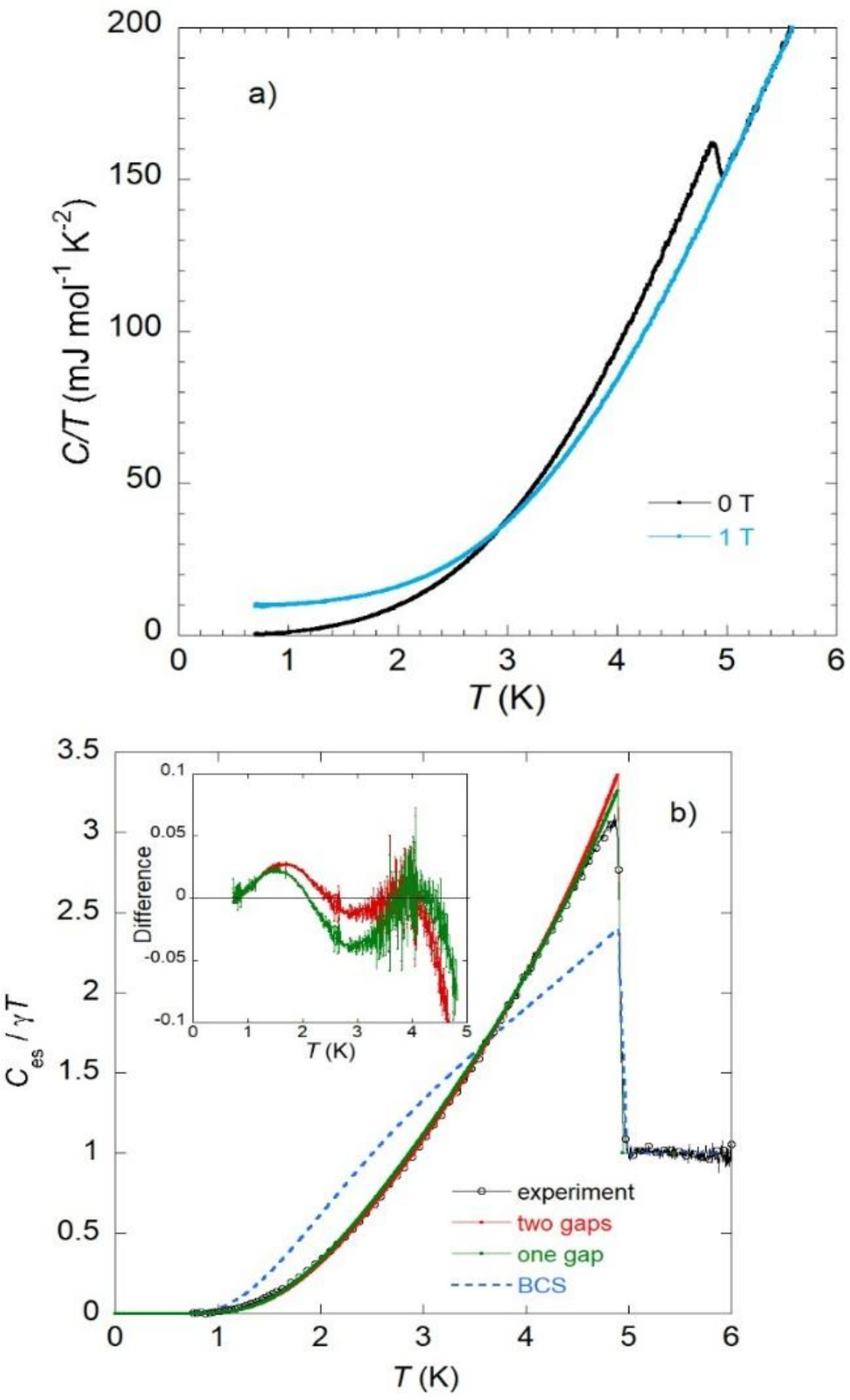

FIG. 2. (a) Total specific heat of $ScAu_2Al$ in zero field (superconducting state, black) and at 1 T (normal state, blue). (b) Normalized superconducting electronic contribution $C_{es}/\gamma_n T = (C_s - C_n)/\gamma_n T + 1$ (black), compared with weak-coupling BCS behavior (blue dashed line), a single-gap strong-coupling α-model fit with $2\Delta/k_BT_c = 4.5$ (green), and a two-gap fit with $2\Delta_S/k_BT_c = 4.1$ and $2\Delta_L/k_BT_c = 4.8$ (red). Inset: residuals between the models and the experimental data.

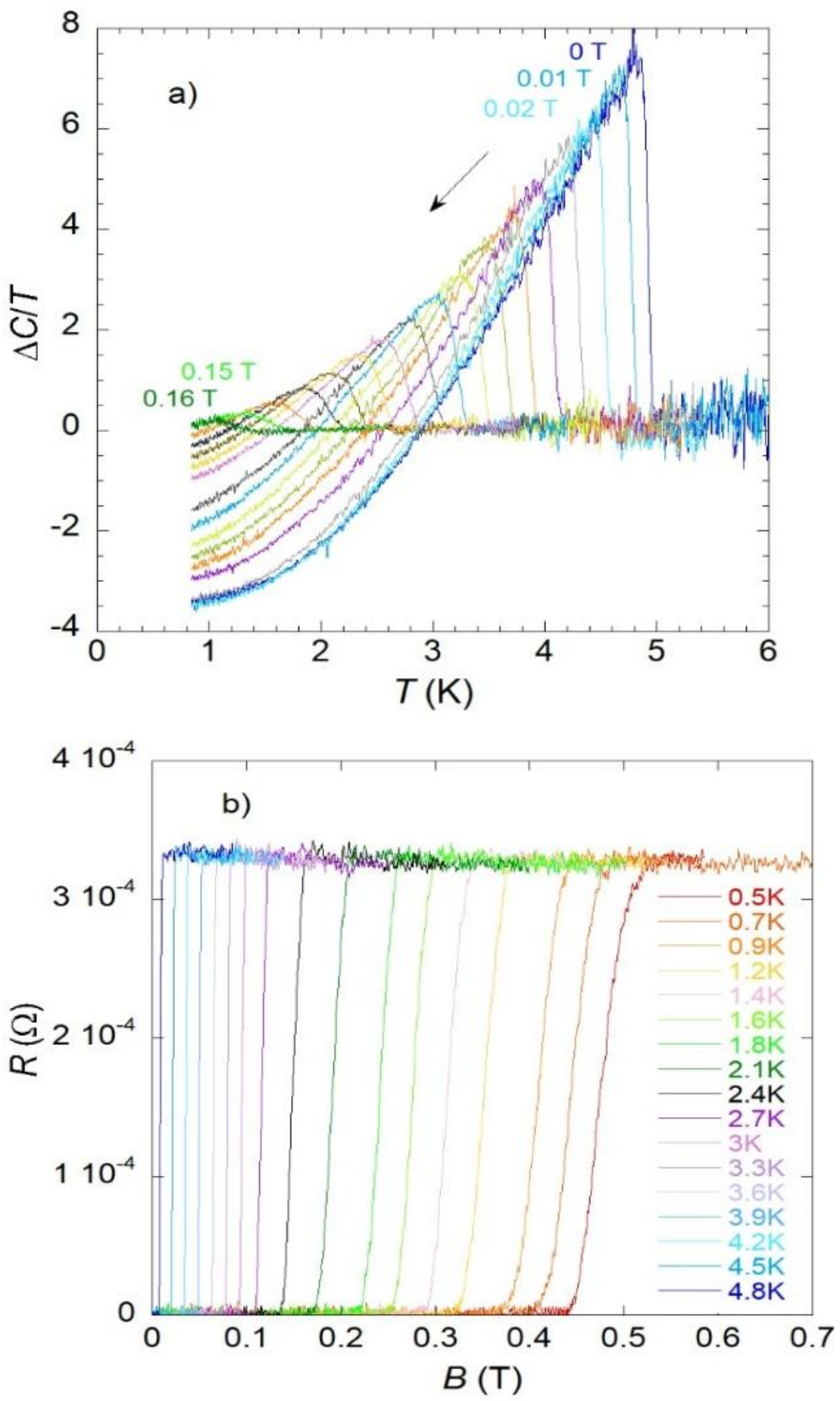

FIG. 3. (a) Superconducting transition in $(C_S - C_n)/T$ at selected magnetic fields. (b) Resistive transitions as a function of magnetic field at selected temperatures.

We analyze the electronic heat capacity using the phenomenological α model [11], in which the gap ratio $2\Delta/k_BT_c$ is allowed to deviate from the weak-coupling BCS value. The model can be extended to two-gap superconductors, as demonstrated for $MgB_2$ [12] and $NbS_2$ [13]. The standard weak-coupling BCS curve clearly fails to reproduce our data. A single-gap strong-coupling fit with $2\Delta/k_BT_c = 4.5$ provides a good description, while a two-gap fit with $2\Delta_S/k_BT_c = 4.1$ and $2\Delta_L/k_BT_c = 4.8$ gives a slightly better agreement. The corresponding partial Sommerfeld weights are $\gamma_S/\gamma_n = 0.30$ and $\gamma_L/\gamma_n = 0.70$. Entropy conservation is satisfied for both fits over the measured temperature range. Although the difference between the two fits is small, the residuals shown in the inset of Fig. 2(b) favor the two-gap description. Additional evidence for multigap superconductivity is discussed below.

In contrast to the previous experimental report of a small superconducting anomaly together with a moderately strong electron-phonon coupling [6], our calorimetric data consistently indicate strong coupling. First-principles calculations including spin-orbit coupling yield a density of states at the Fermi level $N(E_F) \approx 2$ states $eV^{-1}$ per formula unit [7], corresponding to a bare Sommerfeld coefficient $\gamma_{band} = 4.74$ mJ $mol^{-1}$ $K^{-2}$. Using $\gamma_{exp}/\gamma_{band} = 1 + \lambda_{e\text{-}ph}$ with our experimental $\gamma_{exp} \approx 10$ mJ $mol^{-1}$ $K^{-2}$ gives $\lambda_{e\text{-}ph} \approx 1.1$, again placing $ScAu_2Al$ in the strong electron-phonon-coupling regime.

We also investigated the superconducting gap by STM. Unfortunately, the ex situ prepared surface was strongly degraded. Figure S1 of the Supplemental Material [14] shows a tunneling-conductance spectrum measured at 0.5 K, with reduced conductance around zero bias and gap-like peaks near ±0.3 meV. The temperature dependence (Fig. S2) shows that these features disappear near 1 K, far below the bulk $T_c \approx 5$ K. Across tens of locations, the peak energies range from approximately 0.25 to 0.4 meV, with corresponding local $T_c$ values of 1–1.5 K. Bringing the tip closer to the surface and eventually entering the Andreev-reflection point-contact regime shifts the gap-like features to higher energy and increases the local $T_c$ to approximately 2.5 K. In both tunneling and point-contact regimes, the gap-like features are suppressed by fields of approximately 0.5–0.6 T at 0.5 K (Fig. S3). These surface-sensitive measurements therefore probe superconductivity substantially modified from that of the bulk.

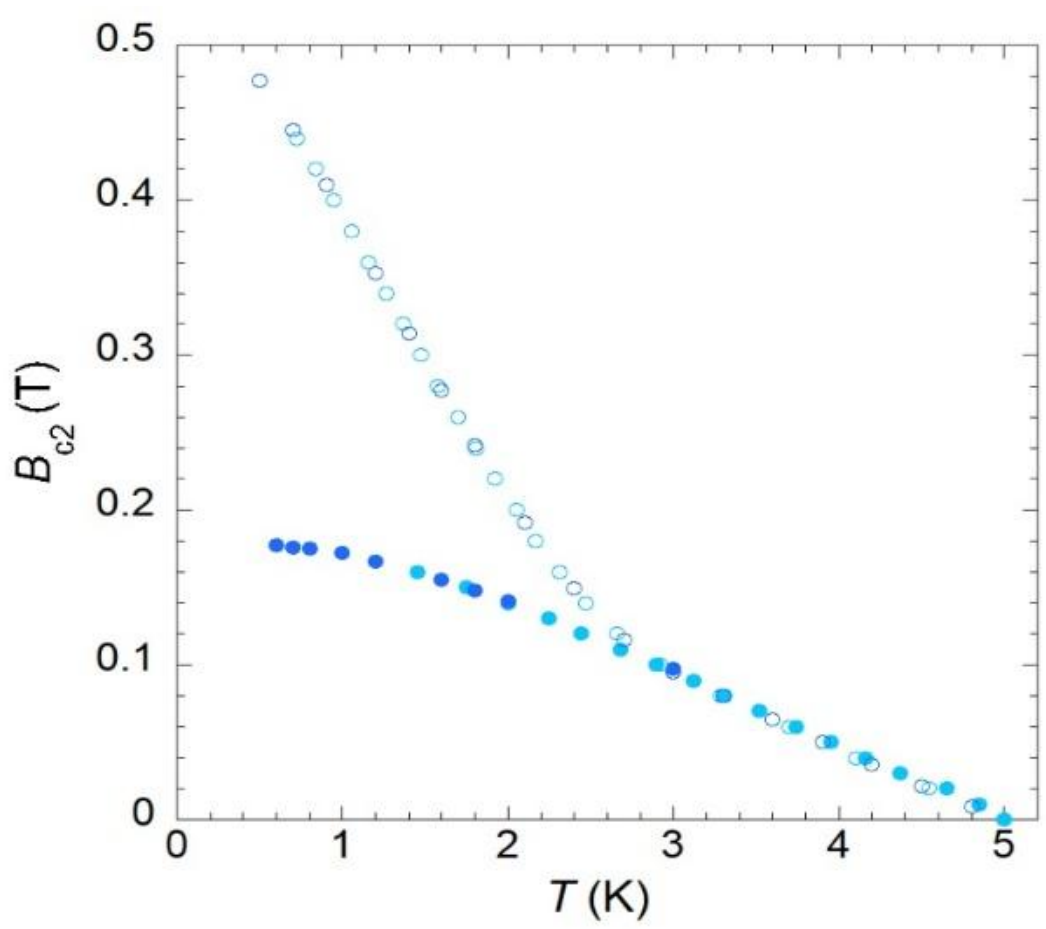

FIG. 4. Temperature dependence of the upper critical field $B_{c2}$ determined from the specific-heat anomaly (closed circles) and from resistive transitions (open circles).

We next examine the magnetic-field dependence of superconductivity calorimetrically. Figure 3(a) presents $\Delta C/T = [C(B) - C(1\ \mathrm{T})]/T$ as a function of temperature for fields from 0 to 0.17 T in steps of 0.01 T. With increasing field, the superconducting anomaly shifts to lower temperature, decreases in magnitude, and broadens, but remains well resolved up to the highest fields. Taking $T_c(B)$ at the midpoint of each transition yields the thermodynamic upper critical field $B_{c2}(T)$, shown by solid circles in Fig. 4. The light-blue points are obtained from the temperature sweeps of Fig. 3(a), whereas the dark-blue points are obtained from field sweeps at fixed temperature. $B_{c2}(T)$ decreases approximately linearly below $T_c$ and saturates at low temperature, with $B_{c2}(0) \approx 0.18$ T. The overall behavior is close to the conventional Werthamer-Helfand-Hohenberg (WHH) form [15], $B_{c2}(0) \approx -0.69 T_{c0}(dB_{c2}/dT)_{Tc}$.

Figure 3(b) shows resistive transitions $R(B)$ measured at temperatures from 0.5 to 4.8 K. The transitions shift systematically to lower fields with increasing temperature and remain relatively narrow, indicating that the electrical transport is not dominated by a random weak-link Josephson network. Remarkably, zero resistance persists to approximately 0.44 T at the lowest temperature, far above the thermodynamic $B_{c2}(0) \approx 0.18$ T. We also measured $R(T)$ at fixed fields in steps of 0.02 T. In both temperature and field sweeps, the resistive critical field was defined at 50% of the normal-state resistance. The resulting $B_{c2}(T)$, shown by open circles in Fig. 4, follows the thermodynamic field near $T_c$ but develops a pronounced upward curvature below approximately 3 K, reaching an extrapolated low-temperature scale of 0.5–0.6 T, about three times the bulk thermodynamic value. Measurements on several pieces cut from the same polycrystalline sample reproduce the thermodynamic $B_{c2}(T)$ essentially exactly, whereas the absolute resistive critical field varies by about 10%.

The heat capacity identifies the bulk thermodynamic upper critical field at which the main volume of the $ScAu_2Al$ grains becomes normal, $B_{c2}(0) \approx 0.18$ T. Nevertheless, after most of the bulk is normal, a connected network of superconducting grain-boundary, interfacial, or near-surface regions can still short-circuit the sample. The resistivity can therefore remain zero up to a substantially higher field. Surface superconductivity may contribute through the third critical field $B_{c3} \approx 1.695\ B_{c2}$ for an ideal surface with field parallel to it [16], while disorder or strain at grain boundaries can further reduce the local mean free path and coherence length. The bulk value $B_{c2}(0) \approx 0.18$ T corresponds to an effective coherence length $\xi \approx 43$ nm. A local coherence length smaller by approximately $\sqrt{3}$ would yield an upper critical field about three times larger. We therefore attribute the resistive field scale primarily to a percolating network of dirtier interfacial regions with reduced coherence length. This interpretation is consistent with the surface-sensitive STM measurements, which reveal gap-like superconducting features persisting up to 0.5–0.6 T at low temperature (Fig. S3).

## IV. THEORETICAL ANALYSIS

As discussed in Ref. [7], the phonon spectrum of $ScAu_2Al$ is unusual, containing a low-frequency, nearly flat acoustic mode that is further softened by spin-orbit coupling. This mode couples strongly to the electrons and contributes $\lambda_1 = 0.453$, approximately 36% of the total calculated electron-phonon coupling $\lambda = 1.253$. The three acoustic branches together contribute about 77% of the total coupling, $\lambda_{1+2+3} = 0.964$, emphasizing the importance of low-energy phonons. The flat mode also produces a non-Debye low-temperature lattice heat capacity, making a conventional polynomial extrapolation of C(T) over a several-kelvin interval unreliable [7].

To further examine the strong-coupling and multigap character of $ScAu_2Al$, we calculated the thermodynamic properties of the superconducting state within isotropic Eliashberg theory, using the implementation described in Refs. [17, 18]. The Eliashberg equations were solved self-consistently using the spin-orbit-coupled spectral function $\alpha^2F(\omega)$ calculated in Ref. [7]. The cutoff frequency was eight times the maximum phonon frequency, 8000 Matsubara frequencies were used, and the convergence criterion was $10^{-15}$ meV. The Coulomb pseudopotential $\mu^*$ was treated as an adjustable parameter and chosen to reproduce the experimental $T_c = 4.95$ K, yielding $\mu^* = 0.145$. The electronic heat capacity was then calculated from the difference between the normal- and superconducting-state free energies and compared with the experimental $C_{es} = C_s - C_n + \gamma_n T$,. The result is shown in Fig. 5.

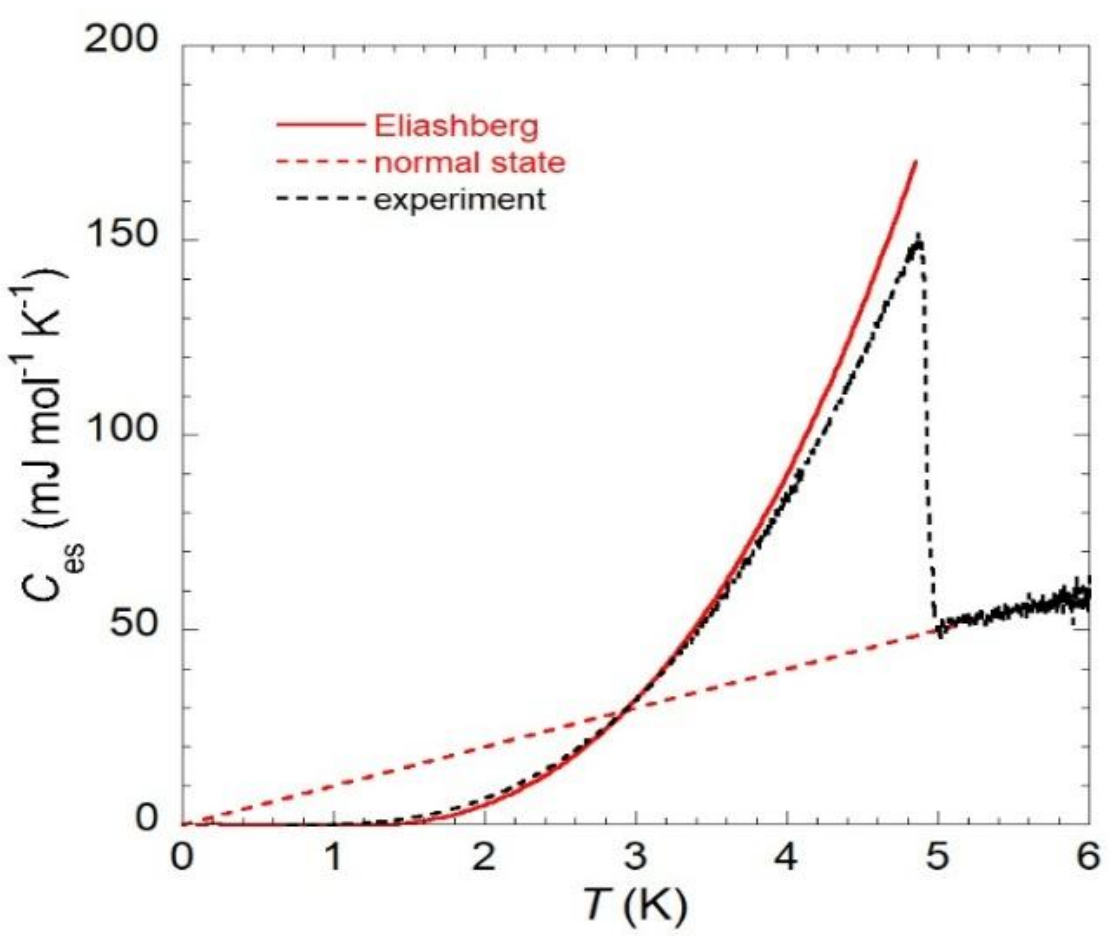

FIG. 5. Electronic heat capacity of $ScAu_2Al$ in the superconducting state. Experimental data are compared with calculations within the isotropic Eliashberg formalism.

The isotropic Eliashberg calculation predicts a large heat-capacity jump $\Delta C/\gamma_n T_c = 2.52$, well above the weak-coupling BCS value of 1.43 and slightly larger than the experimental value of 2.2. However, the isotropic single-gap calculation does not reproduce the detailed temperature dependence of $C_{es}(T)$. To examine this discrepancy, we extended the calculations of Ref. [7] using density-functional theory for superconductors (SCDFT) [19, 20] and the SCTK package [21, 22] including spin-orbit coupling and spin fluctuations yielding $T_c = 4.79$ K. As a result a distribution of superconducting gaps at various temperatures is obtained as shown in Fig. 6, where blue color depicts gaps on the hole-like band and the red color on the electron-like band as described in Ref.[7]. Remarkably, the gap distribution remains distinctly non-single-valued even close to $T_c$, and the spectral weight at smaller $\Delta$ increases with temperature. Gap anisotropy reduces the heat-capacity anomaly near $T_c$ relative to the isotropic result [23], consistent with experiment. At low temperature, where the heat capacity is particularly sensitive to the minimum gap, the isotropic solution lies below the measured $C_{es}(T)$, again consistent with a distribution extending to smaller gap values. Thus, the deviations from the isotropic Eliashberg calculation provide thermodynamic support for the multigap/anisotropic superconducting state predicted by SCDFT.

A second notable feature is the large difference between the thermodynamic and resistive upper critical fields. To examine the bulk field scale, we calculated $B_{c2}(T)$ within isotropic Eliashberg theory [17, 24]. In addition to Eliashberg function $\alpha^2F(\omega)$, this calculation requires the Fermi velocity and electronic scattering time. The calculated average Fermi velocity is $v_F = 6.1 \times 10^5$ m s$^{-1}$ [7]. The scattering time was estimated from the residual resistivity using the Boltzmann transport formalism in the constant-relaxation-time approximation [25]. Electronic eigenvalues were calculated with the full-potential linearized augmented-plane-wave method implemented in WIEN2k [26, 27] on a dense mesh of 200 000 k points, including spin-orbit coupling. The resulting Fermi surface, Fermi velocity, and density of states agree with the pseudopotential calculations of Ref. [7]. At 5 K, the calculated conductivity gives $\sigma/\tau = 7.03 \times 10^{20}$ $\Omega^{-1}$ m$^{-1}$ s$^{-1}$. Together with the measured residual resistivity $\rho_0 = 3.8(5)$ μΩ cm, this yields $\tau \approx 37.5$ fs.

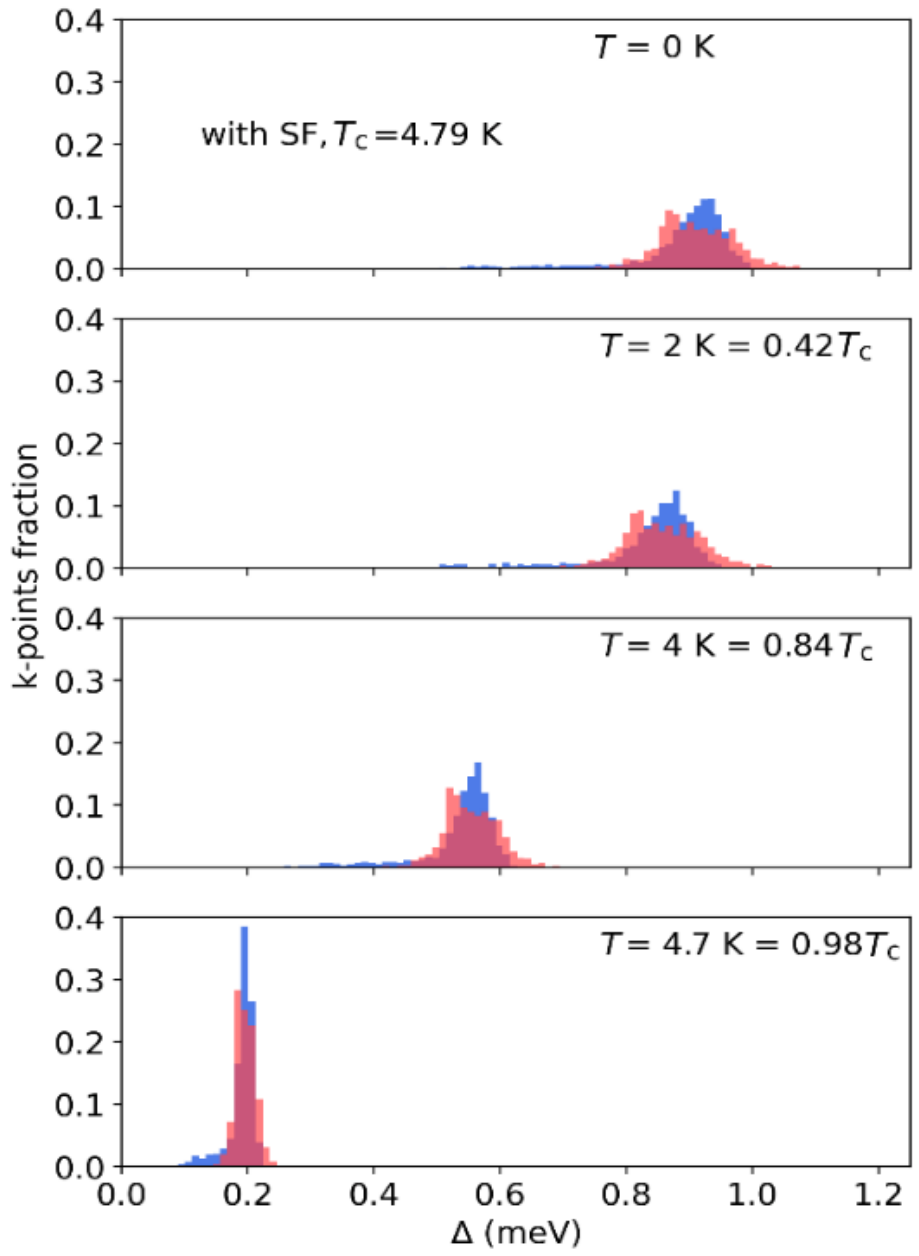

FIG. 6. Temperature evolution of the superconducting-gap distribution in $ScAu_2Al$ obtained from SCDFT calculations. Blue color depicts gaps on the hole-like band and the red color on the electron-like band, details of the calculations are given in Ref. [7].

Figure 7 compares the calculated $B_{c2}(T)$ with the heat-capacity data. The calculated and experimental values agree well above approximately 3.5 K, while the experimental $B_{c2}$ becomes moderately larger at lower temperatures. This discrepancy may reflect multigap effects and anisotropy, including anisotropy of the scattering time, which are not included in the isotropic calculation. Because no additional free parameters are introduced in the $B_{c2}(T)$ calculation, the obtained agreement supports the identification of the calorimetric field scale with the bulk superconducting phase.

The same conclusion follows from a simple coherence-length estimate. The calculated $v_F$ and $\tau$ correspond to a mean free path $\ell = v_F\tau \approx 23$ nm. The clean-limit BCS coherence length, $\xi_0 = \hbar v_F/(\pi\Delta)$, is approximately 130 nm for an average gap $\Delta \approx 1$ meV, placing the sample in the dirty-limit regime. Using $\xi_{dirty}(0) \approx 0.855(\xi_0\ell)^{1/2}$ gives $\xi_{dirty} \approx 46.7$ nm and hence $B_{c2}(0) = \Phi_0/[2\pi\xi_{dirty}^2(0)] \approx 0.15$ T. This value is close to both the isotropic Eliashberg result, $B_{c2}(0) \approx 0.16$ T, and the experimental thermodynamic value, 0.18 T. The consistency of these independent estimates confirms that the heat-capacity measurements probe the bulk upper critical field of the Eliashberg-type superconducting phase in $ScAu_2Al$.

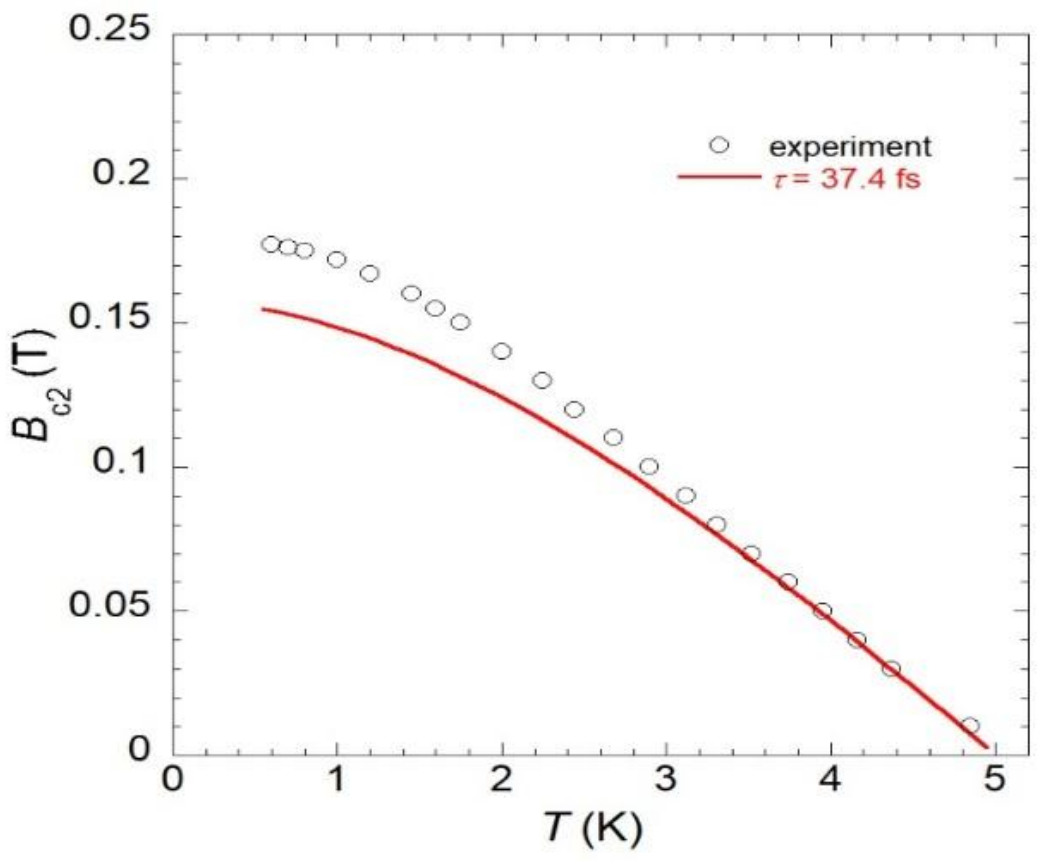

FIG. 7. Upper critical field calculated within the Eliashberg formalism using an electronic scattering time τ = 37 fs obtained from the resistivity analysis. Symbols show the experimental values determined from heat-capacity measurements.

## CONCLUSIONS

We have investigated superconductivity in the Heusler compound $ScAu_2Al$ by high-resolution ac calorimetry, electrical transport, and first-principles-based strong-coupling calculations. The specific heat reveals a sharp superconducting transition at $T_c \approx 4.95$ K and a large normalized jump $\Delta C/\gamma_n T_c \approx 2.2$, establishing strong-coupling superconductivity. The low-temperature electronic specific heat is well described by a strong-coupling α model; a two-gap description provides a slightly better fit than a single-gap model and is consistent with first-principles predictions of multiband superconductivity. The measured Sommerfeld enhancement gives $\lambda_{e\text{-}ph} \approx 1.1$, in good agreement with the calculated Eliashberg coupling $\lambda \approx 1.25$.

The bulk upper critical field determined calorimetrically follows a conventional WHH-like temperature dependence with $B_{c2}(0) \approx 0.18$ T. Isotropic Eliashberg calculations yield a similar field scale, and the remaining low-temperature difference is plausibly associated with multiband and anisotropy effects. In contrast, the resistive critical field reaches approximately 0.5–0.6 T at low temperature. We attribute this discrepancy to a percolating superconducting network associated with grain boundaries, interfaces, and locally dirtier regions of the polycrystalline sample, where a reduced coherence length allows superconductivity to persist above the bulk $B_{c2}$.Taken together, the calorimetric and theoretical results establish $ScAu_2Al$ as a strong-coupling, multigap Heusler superconductor.

## ACKNOWLEDGMENTS

The work in Košice was supported by the Slovak Research and Development Agency under Contracts No. APVV-23-0624 and No. APVV-23-0564 and by the Scientific Grant Agency project VEGA 2/0073/24. The work at AGH University was supported by the National Science Centre (Poland), Project No. 2025/59/B/ST3/01621, and by the Polish high-performance computing infrastructure PLGrid (HPC Center: ACK Cyfronet AGH), computational grant No. PLG/2024/017305. Research performed at Gdańsk University of Technology was supported by the National Science Centre (Poland), Project No. 2022/45/B/ST5/03916.

## SUPPLEMENTARY INFORMATION

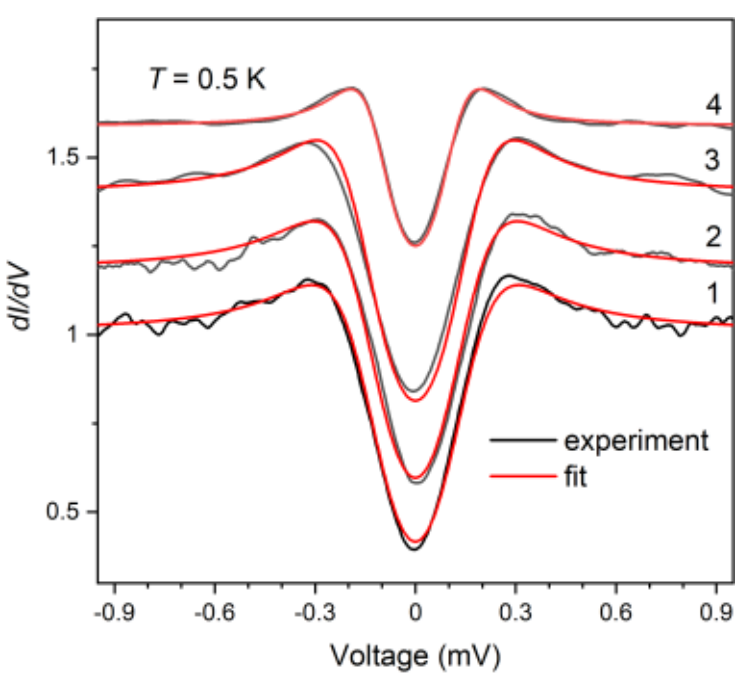

FIG. S1. Representative STM tunneling spectra.

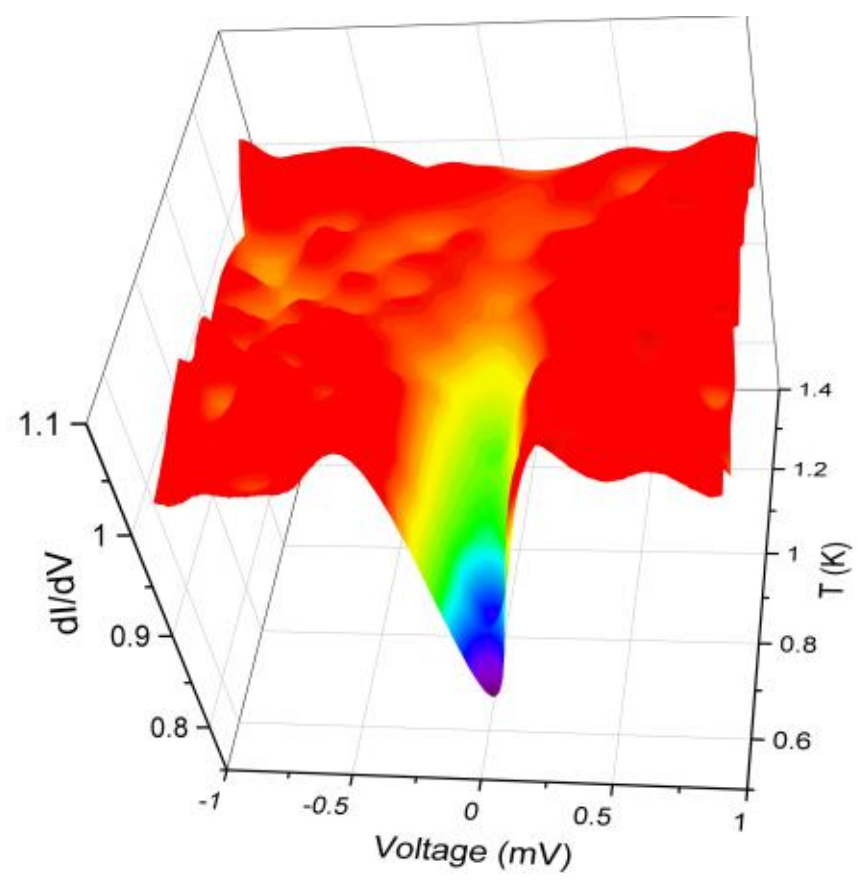

FIG. S2. Temperature evolution of the tunneling spectrum.

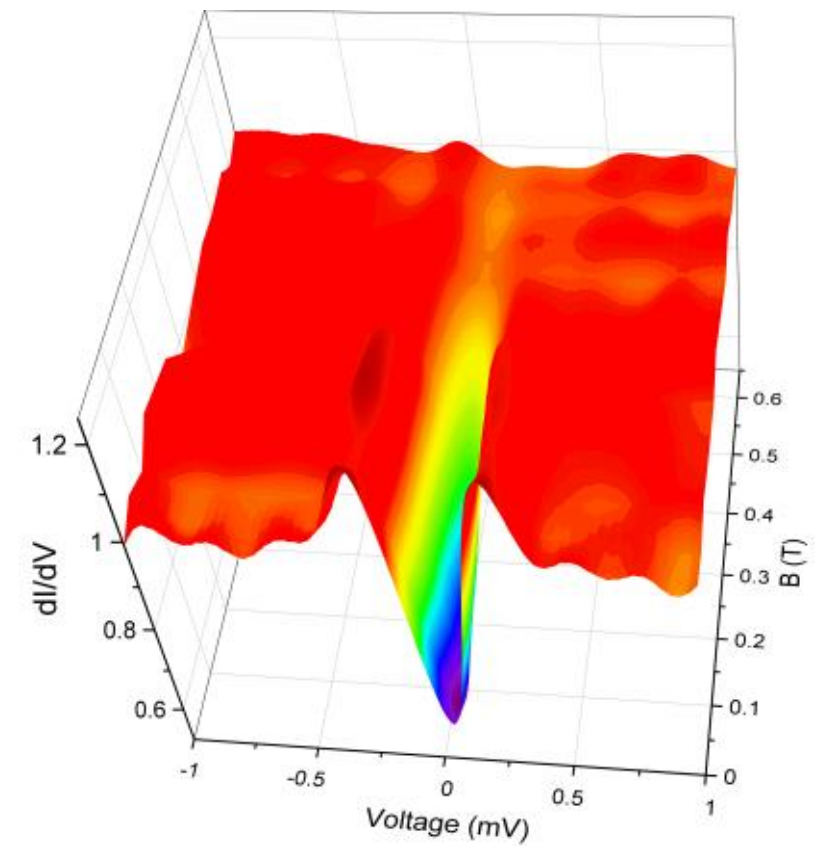

FIG. S3. Magnetic-field evolution of the tunneling spectrum taken at 0.5 K.